# Ultrafast optical nanoscopy of carrier dynamics in silicon nanowires

Jingang Li[1]†, Rundi Yang[1]†, Yoonsoo Rho[1,2]*, Penghong Ci[3,4,5], Matthew Eliceiri[1], Hee K. Park[6], Junqiao Wu[3,4], and Costas P. Grigoropoulos[1]*

[1] Laser Thermal Laboratory, Department of Mechanical Engineering, University of California, Berkeley, California 94720, USA

[2] Physical & Life Sciences and NIF & Photon Sciences, Lawrence Livermore National Laboratory, Livermore, California, 94550, USA

[3] Department of Materials Science and Engineering, University of California, Berkeley, California 94720, USA

[4] Materials Sciences Division, Lawrence Berkeley National Laboratory, Berkeley, California 94720, USA

[5] Institute for Advanced Study, Shenzhen University, Shenzhen 518060, China

[6] Laser Prismatics, LLC, San Jose, USA

† These authors contributed equally to this work.

**Correspondence to:** cgrigoro@berkeley.edu  (C.P.G.), hirho@berkeley.edu (Y.R.)

## ABSTRACT

Carrier distribution and dynamics in semiconductor materials often govern their physical properties that are critical to functionalities and performance in industrial applications. The continued miniaturization of electronic and photonic devices calls for tools to probe carrier behavior in semiconductors simultaneously at the picosecond time and nanometer length scales. Here, we report pump-probe optical nanoscopy in the visible-near-infrared spectral region to characterize the carrier dynamics in silicon nanostructures. By coupling experiments with the point-dipole model, we resolve the size-dependent photoexcited carrier lifetime in individual silicon nanowires. We further demonstrate local carrier decay time mapping in silicon nanostructures with a sub-50 nm spatial resolution. Our study enables the nanoimaging of ultrafast carrier kinetics, which will find promising applications in the future design of a broad range of electronic, photonic, and optoelectronic devices.

## TEASER

A visible-near-infrared pump-probe nanoscopy is reported to probe nanoscale carrier dynamics in silicon nanostructures.

# INTRODUCTION

Semiconductor technology has stimulated the exponential progress of microelectronics in the last half-century. Silicon-based complementary metal-oxide-semiconductor integrated circuits form the backbone of modern technology (*1, 2*). In addition to the prevalence in electronic devices, silicon nanostructures are widely exploited in photonics for versatile control of light propagation, detection, and modulation (*3-5*). Recently, emerging low-dimensional materials, particularly silicon nanowires (SiNWs), have also opened new possibilities for broad applications in electronics, photonics, photovoltaics, and photoelectrochemistry (*6-9*).

For most semiconductor materials, free-carrier concentration determines their electrical and optical properties, such as conductivity, refractive index, and absorption coefficients (*10-12*). In addition, the performance of transistors and semiconductor devices primarily depends on the spatial and temporal carrier distribution (*13-16*). Therefore, resolving the charge transport and carrier dynamics in semiconductor nanostructures is essential for designing future electronic and photonic devices.

Time-resolved terahertz spectroscopy has been widely applied to reveal carrier dynamics in bulk and nanostructured semiconductors (*17, 18*). In addition, pump-probe microscopy was developed to probe the site-specific carrier transport and recombination (*19-21*). However, the spatial resolution of this microscopic method is limited by optical diffraction. Alternatively, tip-based scattering-type scanning near-field optical microscopy (s-SNOM) is capable of mapping free-carrier distribution in semiconductor nanostructures with a nanoscale resolution (*22, 23*). Ultrafast infrared-terahertz nano-spectroscopy was further developed through the integration of s-SNOM and pump-probe optics. However, given the limited photon energy, the efforts have been primarily focused on studying carrier dynamics in narrow bandgap semiconductors, such as InAs (*24, 25*) and $Hg_{1-x}Cd_xTe$ (*26*), or graphene plasmon (*27*).

Here, we report near-field ultrafast optical nanoscopy in the visible-near-infrared spectral region to access the carrier dynamics in silicon, one of the most prevalent materials in current semiconductor technology. Our pump beam has a wavelength of 400 nm (3.1 eV), which is sufficient to excite carriers in common optoelectronic semiconductors, including silicon (bandgap of 1.12 eV) and GaAs (bandgap of 1.42 eV). By combining ultrafast nanoscale measurements and theoretical modeling, we unravel the local photocarrier recombination dynamics in silicon nanowires. Moreover, we demonstrate the spatial mapping of carrier lifetime in silicon with a sub-50 nm resolution. Our results provide the capability to probe carrier behaviors in nanoscale materials and devices, which is of great significance to understanding the optoelectronic properties and practical functionality of semiconductor nanostructures.

# RESULTS

Fig. 1A shows the schematic of our pump-probe near-field optical nanoscopy. A 400 nm pump beam and an 800 nm probe beam with a controlled delay time $\Delta t$ are directed to an atomic force microscope (AFM) tip with an oscillating frequency $\Omega$. (see Materials and Methods and Fig. S1 for more details on the setup). The backscattered probe beam is detected and demodulated at the third harmonic of the tapping frequency to suppress the background noise (*28, 29*). The samples in this study are chemically synthesized SiNWs (see Inset in Fig. 1I). Fig. 1B and Figs.1C-G show the topography of a SiNW and the time-resolved s-SNOM imaging at different time delays, respectively. The high spatial overlap of the height and s-SNOM profiles indicates the successful detection of near-field signals (Fig. 1H). The time-dependent snapshots distinctly exhibit a dynamic change in the intensity of scattered light (Fig. 1J). Specifically, the s-SNOM signal amplitude undergoes an intense increase (Fig. 1D), followed by a decay process (Figs. 1E and F) and a slow rise at longer delay times (Fig. 1G). This result can also be quantitatively visualized in the transient curve presented in Fig. 1I.

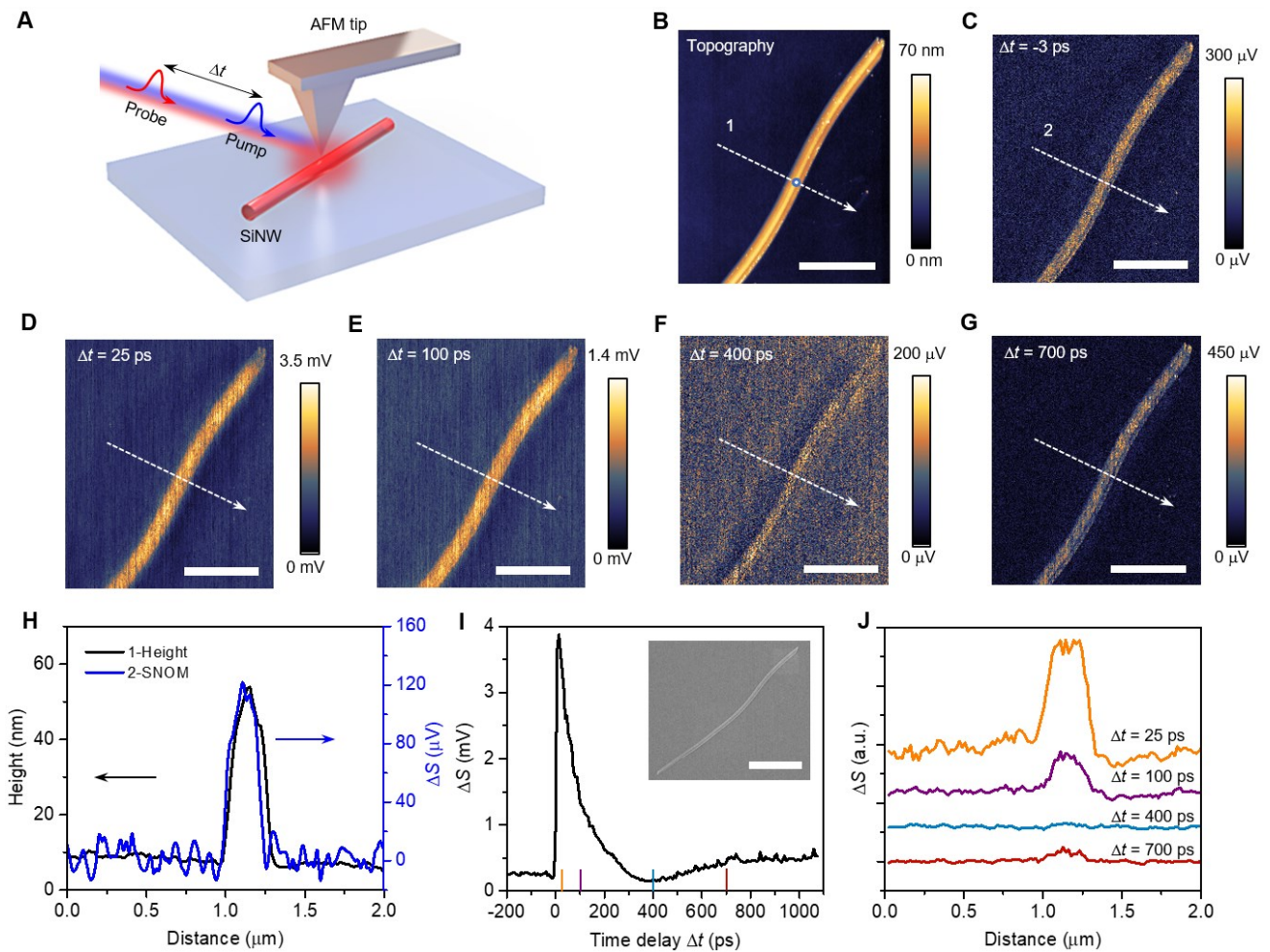

**Fig. 1. Time-resolved near-field imaging of a SiNW.** (**A**) Schematic of the pump-probe optical nanoscopy setup. The wavelengths of the pump and probe beams are 400 and 800 nm, respectively. (**B**) AFM topography of a SiNW.

(**C** to **G**) Time-resolved s-SNOM imaging of the SiNW with a pump-probe delay time of (**C**) -3 ps, (**D**) 25 ps, (**E**) 100 ps, (**F**) 400 ps, and (**G**) 700 ps. (**H**) Height and s-SNOM profiles along the dashed lines in (**B**) and (**C**). (**I**) Transient s-SNOM signal as a function of delay time measured at the SiNW (marked by the dot in (**B**)). Inset shows the SEM image of the SiNW. (**J**) s-SNOM line profiles across the SiNW at different pump-probe delay times in (**D** to **G**). Scale bars: (**B** to **G**) 1 μm, Inset in (**I**): 2 μm.

To understand the evolution of the near-field scattering signal over the increasing delay time, we adopt a point-dipole model to analyze the scattered light (see Note S1) (*23, 30, 31*). Briefly, the scattered field **E**$_{sc}$ by the tip is derived from the dipolar near-field interaction between the AFM tip and the sample with the dielectric function $\varepsilon$, yielding (*30, 31*)

$$\mathbf{E}_{sc} \propto A e^{i\varphi} \mathbf{E}_{in} = \frac{\alpha(1+\beta)}{1-\frac{\alpha\beta}{16\pi(a+z)^3}} \mathbf{E}_{in} \quad (1)$$

Here the measured scattering intensity $S \propto A^2$, $\varphi$ is the phase of scattered light, $a$ is the AFM tip radius, $z$ is the tip-sample distance, $\alpha = 4\pi a^3$ is the polarizability of the tip, and $\beta = (\varepsilon-1)/(\varepsilon+1)$, respectively (Fig. 2A). The dielectric function $\varepsilon$ is further calculated as a function of free carrier density $N$ based on the Drude model (Fig. S2) (*32*)

$$\varepsilon(\omega, N) = \varepsilon_\infty - \frac{\omega_p(N)^2}{\omega(\omega+i\Gamma)} \quad (2)$$

where $\varepsilon_\infty = 11.7$ for silicon, $\omega_p$ is the plasma frequency and a function of the carrier concentration $N$ (see Note S2), and $\Gamma$ is the damping rate. In addition, the effect of temperature is considered by the Jellison-Modine model (see Fig. S3 and Note S2 for more details) (*33*).

Fig. 2B shows the calculated s-SNOM intensity with an increasing free-carrier concentration $N$ from $10^{18}$ to $10^{22}$ cm$^{-3}$ at the wavelength of 800 nm. As $N$ increases, the calculated s-SNOM signal exhibits a minimum at $N \sim 6 \times 10^{19}$ cm$^{-3}$ and then reaches its maximum at $N \sim 1.2 \times 10^{20}$ cm$^{-3}$. This behavior arises from the resonant near-field interaction between the AFM probing tip and the free carriers in SiNWs (*30*). The experimental transient s-SNOM signal can be well fitted by the point-dipole model assuming a biexponential decay of photoexcited carriers, i.e., $N(\Delta t) = N(0) \times (A_1 e^{-\frac{\Delta t}{t_1}} + A_2 e^{-\frac{\Delta t}{t_2}})$ (Fig. 2C). The biexponential kinetics ($t_1$, $t_2$) are attributed to the carrier recombination and diffusion (*34*), and the average carrier lifetime $t_{avg}$ can be obtained by $t_{avg} = (A_1 + A_2)/(A_1/t_1 + A_2/t_2)$. We exclude the effect of laser heating as the temperature increase is negligible with respect to the s-SNOM amplitude (Note S3 and Fig. S4). In addition, the variation of tip-sample distance due to thermal expansion of SiNWs only leads to slight changes in the scattered light intensity (Note S4 and Fig. S5). These results verify the dominant role of carrier density in the transient s-SNOM measurements. We further validate the point-dipole model by measuring the same SiNW with different pump power (Fig. 2D), where two curves can be well fitted with different initial carrier densities and the same decay kinetics (Table S1).

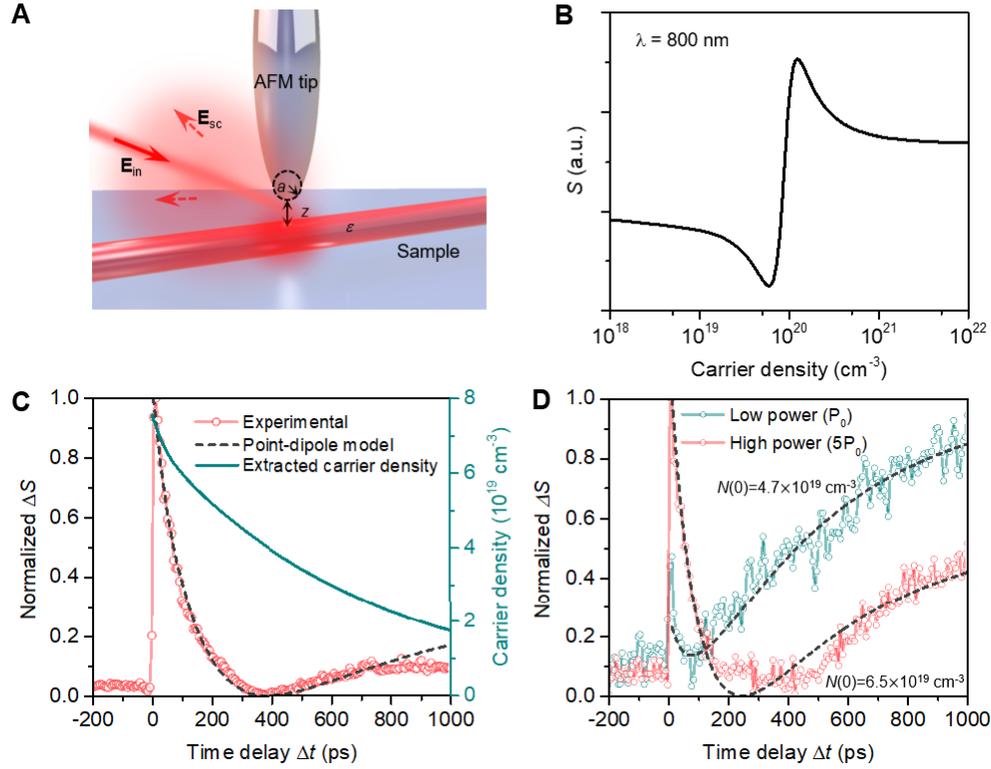

**Fig. 2. Point-dipole modeling.** (**A**) Schematic of the point-dipole model. $a$, $z$, and $\varepsilon$ are the AFM tip radius, tip-sample distance, and the permittivity of the sample, respectively. (**B**) Modeled s-SNOM intensity as a function of carrier density in SiNWs. (**C**) Experimental transient s-SNOM signal, fitting curve with the point-dipole model, and the extracted carrier density in SiNW as a function of pump-probe delay time. (**D**) Experimental transient s-SNOM scans and fitting curves under different pump excitation powers.

We then apply pump-probe s-SNOM to investigate the carrier dynamics in individual SiNWs with varying geometries. Fig. 3A shows the transient s-SNOM signals measured in different nanowires with various widths. The corresponding scanning electron microscope (SEM) images are shown in Fig. 3B. All experimental curves are well fitted with the point-dipole model, and the decay times are summarized in Fig. 3C. The carrier lifetime shows a linear increase with the increasing size of the SiNWs, indicating that surface recombination dominates in semiconductor nanowires (*35*). The surface recombination velocity (SRV) can be calculated from carrier lifetime $t_{avg}$ as SRV = $d/4t_{avg}$, where $d$ is the SiNW width (*36*). The linear fitting gives a surface recombination velocity of $2.2 \times 10^4$ cm/s, which is consistent with previous reports (*34, 37*). These results validated that our tool and model are effective to probe the carrier lifetime in semiconductor nanomaterials.

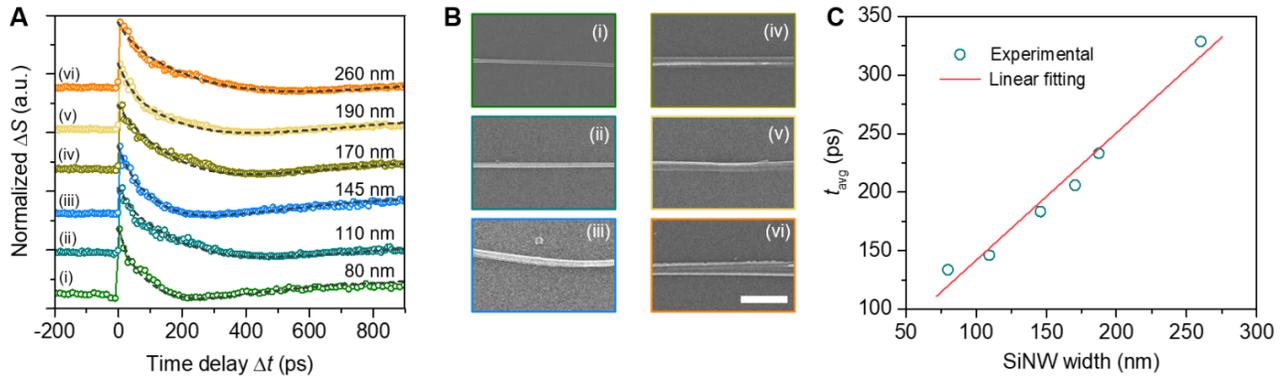

**Fig. 3. Size-dependent carrier dynamics.** (**A**) Experimental transient s-SNOM scans measured from SiNWs with different widths (i) to (vi). The grey dashed curves present the fitting results by the point-dipole model. (**B**) Corresponding SEM images of the SiNWs in (A). (**C**) Size-dependent average decay time and the linear fitting. Scale bar: (B) 1 μm.

We further explore the capability of near-field ultrafast nanoscopy to probe the spatially resolved carrier dynamics with a nanoscale resolution, which cannot be accessed with conventional pump-probe microscopy. A nonuniform silicon nanowire is selected as the test sample (Fig. 4A). Transient s-SNOM signals measured at different locations (P1-P4 in Fig. 4A) show distinct temporal evolution (Fig. 4B), corresponding to carrier lifetimes of 460.3, 379.9, 294.6, and 338.0 ps, respectively. The time-resolved near-field images also exhibit spatially nonuniform dynamics of the nanostructure (Fig. S6). Fig. 4C presents the time-resolved s-SNOM map along the nanowire (dashed line in Fig. 4A). The carrier lifetime extracted from the spatiotemporal mapping is strongly correlated to the topography, as evidenced by the qualitatively similar profiles (Fig. 4D). The nonuniform carrier lifetime is mediated by the spatial heterogeneity in the silicon nanostructure (*38*). The mapping of carrier dynamics at a spatial resolution of 35 nm is demonstrated. These results further indicate that our near-field ultrafast nanoscopy can probe the local doping, defects, and barriers at the nanoscale in semiconductor materials and devices.

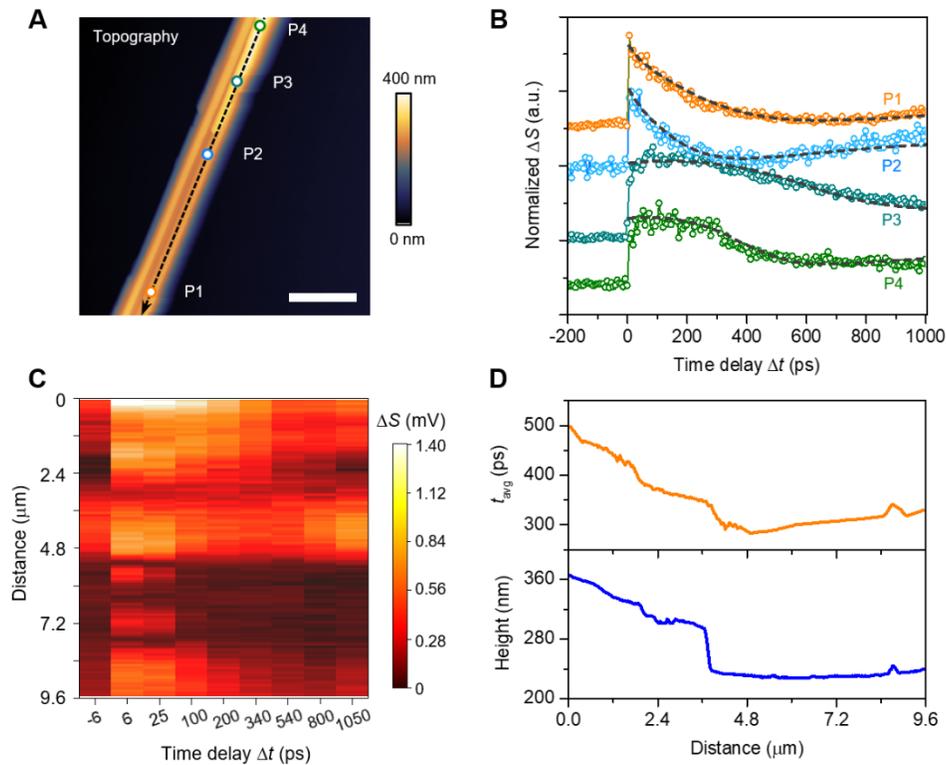

**Fig. 4. Probing spatially resolved carrier dynamics at the nanoscale.** (**A**) AFM topography of a nonuniform silicon nanowire structure. (**B**) Experimental transient s-SNOM scans and fitting curves measured at different locations, as marked in (A). (**C**) Time-resolved s-SNOM mapping along the dashed line in (A). (**D**) Corresponding decay time and height profiles with a spatial resolution of 35 nm. Scale bar: (A) 2 μm.

## DISCUSSION

We have demonstrated visible-near-infrared pump-probe near-field nanoscopy as a noninvasive tool to examine the carrier dynamics in silicon nanostructures. The combination of ultrafast optics and near-field imaging permits the investigation of carrier kinetics with both high temporal (picosecond) and spatial (sub-50 nm) resolutions. In addition to silicon, the use of a visible pump beam enables the investigation of photocarrier dynamics in a wide range of semiconductor materials that are commonly used in practical applications, including germanium and III-V materials. This capability is valuable for the characterization and optimization of functional optoelectronic devices. While this work analyzed the carrier dynamics in SiNWs as a demonstration, the proposed pump-probe near-field nanoscopy can serve as a versatile and general optical diagnostic platform to study nanomaterials, including two-dimensional materials (*39, 40*) and quantum dots (*41*). It also provides a promising tool to study other non-equilibrium thermodynamic phenomena in nanomaterial systems, including phase transitions (*42-44*), energy and charge transfer (*45*), and phonon propagation (*46*).

## MATERIALS AND METHODS

**Materials and Characterizations.**

SiNWs in this study were synthesized via an electroless etching method (*47*). All SEM images are taken with a FEI Quanta 650 SEM.

**Experimental setup.**

The detailed schematic of the experimental setup is shown in Fig. S1. The setup is developed based on a commercial near-field optical microscopy system from Molecular Vista. The platinum-coated AFM tip (Molecular Vista) has a tapping frequency $\Omega$ at ~250 kHz and an apex radius of ~25 nm. An 800 nm femtosecond laser beam (Spectra Physics) is split by a beam splitter into a probe beam and a pump beam. The pump beam is frequency doubled by a nonlinear crystal, beta barium borate (Eksma Optics), and then amplitude-modulated by an acoustic-optic modulator (AOM). The modulation frequency is set to be the third harmonic of the tapping frequency of the AFM tip. The probe beam passes through a mechanical delay stage (Thorlabs). Both beams are merged and directed to the AFM tip via a parabolic mirror. The typical optical intensities of pump and probe beams before entering the AFM chamber are ~1 mW and ~0.5 mW, respectively. The scattered probe light is then collected by the parabolic mirror and redirected to an avalanche photodiode detector (Thorlabs) after it is interfered with a reference beam from the reference mirror, per the scheme known as homodyne detection (*28*). The signal from the avalanche photodiode detector (Thorlabs) is sent to a lock-in amplifier for signal demodulation at the same modulation frequency of the pump beam, i.e., the third harmonic of the tapping frequency, to suppress the background noise. A long-pass filter (Thorlabs) is placed before the detector to block the scattered pump beam.

## SUPPLEMENTARY MATERIALS

Supplementary material for this article is available at https://science.org/doi/XXX

**Acknowledgments:** We thank Zhengliang Su and Qiye Zheng for their assistance in the experimental setup. **Funding:** C.P.G. acknowledges the financial support from Laser Prismatics under the DOE SBIR Phase 2 grant DE-SC0018461. Part of this work by Y.R. was performed under the auspices of the U.S. Department of Energy by Lawrence Livermore National Laboratory under Contract DE-AC52-07NA27344. The nanowire preparation was supported by the U.S. Department of Energy, Office of Science, Office of Basic Energy Sciences, Materials Sciences and Engineering Division under Contract No. DE-AC02-05-CH11231 (EMAT program KC1201). **Author contributions:** J.L., Y.R., and C.P.G. conceived the idea and planned the experimental work. J.L., R.Y., and Y.R. worked on the experiments and analyzed the data. P.C. and J.W. synthesized the silicon nanowires. M.E. assisted in the experiments. H.K.P. contributed to the experimental design and apparatus construction. J.L. wrote the paper with inputs from all authors. C.P.G. supervised the project. **Competing interests:** The authors declare no competing interests. **Data and materials availability:** All data needed to evaluate the conclusions in the paper are present in the paper and/or the Supplementary Materials.

# Supplementary Materials for

**Supplementary Notes**

**Note S1. Point-dipole model**

The point-dipole model provides a good qualitative description of the near-field interaction in scattering-type scanning near-field optical microscopy (s-SNOM) (*30, 31, 48*). The tip is reduced to a small sphere with a point dipole at its center. The scattered field is calculated based on the dipolar near-field coupling between the tip and sample. Briefly, the scattered field $\mathbf{E}_{sc}$ can be written as

$$\mathbf{E}_{sc} \propto Ae^{i\varphi}\mathbf{E}_{in} = \frac{\alpha(1+\beta)}{1-\frac{\alpha\beta}{16\pi(a+z)^3}}\mathbf{E}_{in} \quad (S1)$$

Here the measured scattering intensity $S \propto A^2$, $\varphi$ is the phase of scattered light, $a$ is the AFM tip radius, $z$ is the tip-sample distance, $\alpha = 4\pi a^3$ is the polarizability of the tip, and $\beta = (\varepsilon - 1)/(\varepsilon + 1)$, respectively. In our case, $a = 20$ nm is used. $\varepsilon$ is calculated based on the Drude model and the Jellison-Modine model to account for the effect of carrier density and temperature, respectively, which will be discussed in Note S2. The time-resolved s-SNOM amplitude is calculated based on the biexponential decay of photocarriers

$$N(\Delta t) = N(0) \times (A_1 e^{-\frac{\Delta t}{t_1}} + A_2 e^{-\frac{\Delta t}{t_2}}) \quad (S2)$$

where $N(0)$ is the initial carrier density right after the pump excitation, $\Delta t$ is the pump-probe time delay, $A_1$, $A_2$, $t_1$, and $t_2$ are fitting parameters with $A_1 + A_2 = 1$. The fitting of experimental curved is conducted in MATLAB based on a nonlinear least-squares algorithm to obtain the decay time parameters.

**Note S2. Modeling of the dielectric function $\varepsilon$**

The dielectric function $\varepsilon$ is modeled as a function of carrier density $N$ by the Drude model (*32*)

$$\varepsilon(\omega) = \varepsilon_\infty - \frac{\omega_p^2}{\omega(\omega+i\Gamma)} \quad (S3)$$

here $\varepsilon_\infty = 11.7$ for silicon, $\Gamma$ is the damping rate, $\omega_p = \frac{Ne^2}{\varepsilon_0 m}$ is the plasma frequency where $e$ is the elementary charge, $\varepsilon_0$ is the free-space permittivity, and $m$ is the effective carrier mass. $\Gamma$ is calculated by $\Gamma = \frac{e}{m\mu}$, where $\mu$ is the carrier mobility. $\mu$ is determined as a function of $N$ based on the empirical equations proposed by Arora and coworkers (*49*).

In addition, the effect of temperature is considered by the Jellison-Modine model as (*33*)

$$n(\omega, T) = n_0(\omega) + a(\omega)T \quad (S4)$$

$$k(\omega, T) = k_0(\omega)\exp\left(\frac{T}{T_0(\omega)}\right) \tag{S5}$$

where $\omega_g$ =3.648 eV, $n_0(\omega) = \sqrt{4.565 + 97.3/(\omega_g^2 - \omega^2)}$, $a(\omega) = [-1.864 + 53.94/(\omega_g^2 - \omega^2)] \times 10^{-4}$, $k_0(\omega) = -0.0805 + \exp[-3.1893 + 7.946/(\omega_g^2 - \omega^2)]$, and $T_0(\omega) = 369.9 - \exp[-12.92 + 5.509\omega]$.

**Note S3. Exclusion of the laser heating effect**

The pump excitation of photocarriers is accompanied by the laser heating effect, which causes the increase in lattice temperature. This effect of temperature rise is considered by the Jellison-Modine model (Equations S4 and S5), and the obtained dielectric function is used to calculate s-SNOM amplitude based on the point-dipole model. We model the s-SNOM amplitude change for a temperature range of 300-800 K. Only a slight increase of 1.3% is observed (Fig. S6). In practice, the laser heating is minor, and the temperature is likely to be lower than the simulated range. Thus, we ignore the effects of temperature rise in our analysis.

**Note S4. Exclusion of the thermal expansion effect**

The temperature increase can also lead to the thermal expansion of the sample, thus reducing the tip-sample distance. This effect is directly considered in the point-dipole model by sweeping $d$. Considering the thermal expansion of silicon to be $2.6 \times 10^{-6}\ K^{-1}$, the change in the tip-sample distance is expected to be very small and less than 0.2 nm. This change only leads to a slight change of s-SNOM amplitude of 1.6% (Fig. S6), which indicates that thermal expansion can also be neglected.

**Supplementary Tables and Figures**

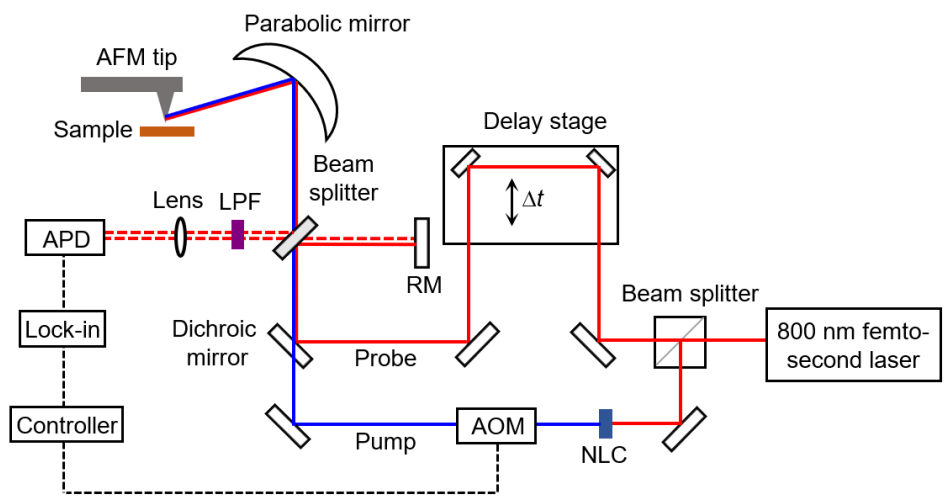

**Fig. S1. Experimental setup of pump-probe s-SNOM.** NLC: non-linear crystal; AOM: acoustic-optic modulator; APD: avalanche photodiode; LPF: long-pass filter; RM: reference mirror. Detailed description of the setup can be found in Materials and Methods section.

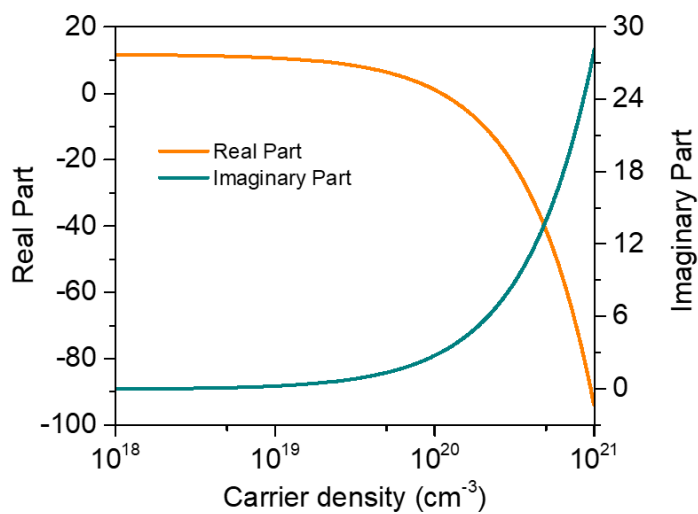

**Fig. S2. Dielectric function of silicon as a function of carrier density calculated by the Drude model.**

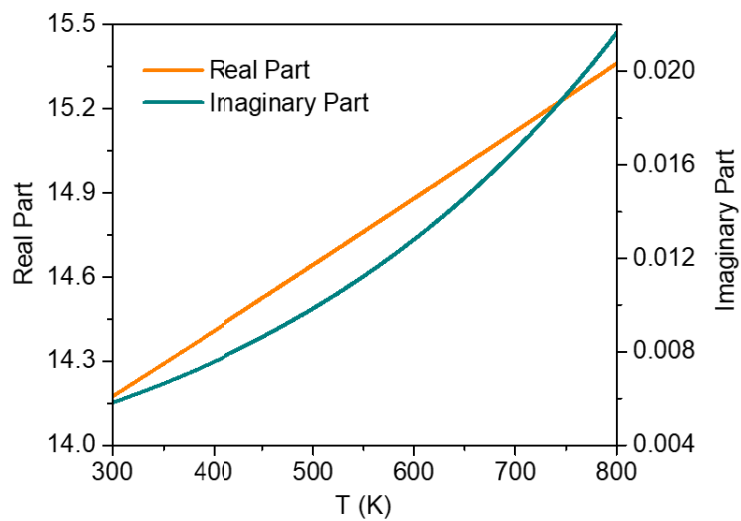

**Fig. S3. Dielectric function of silicon as a function of temperature calculated by the Jellison-Modine model.**

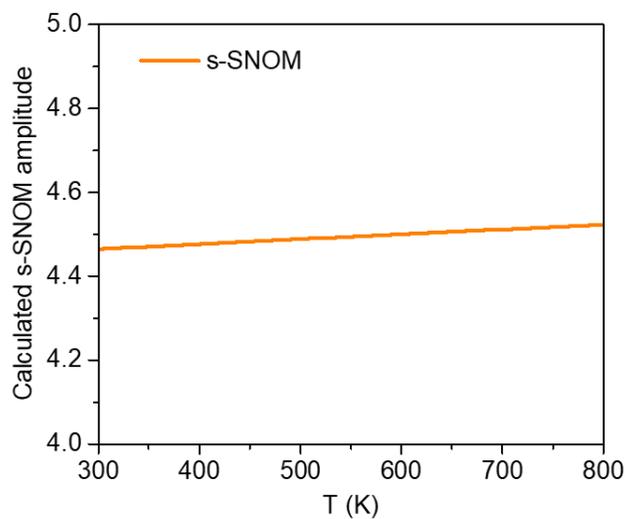

**Fig. S4. Calculated s-SNOM amplitude when the temperature increases from 300 K to 800 K.**

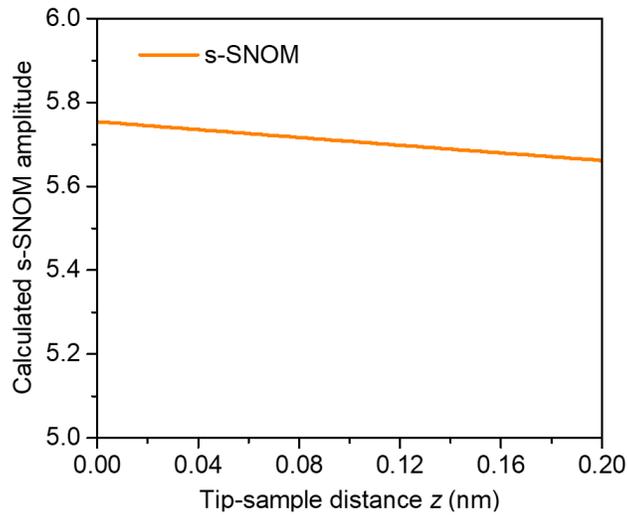

**Fig. S5. Calculated s-SNOM amplitude when the-tip sample distance increases from 0 nm to 0.2 nm.**

**Table S1. Fitting parameters in Fig. 2.** ($N(t) = N(0) \times (A_1 e^{-\frac{t}{t_1}} + A_2 e^{-\frac{t}{t_2}})$, $A_1 + A_2 = 1$)

| Data | $N(0)$ (×10$^{19}$ cm$^{-3}$) | $A_1$ | $t_1$ (ps) | $A_2$ | $t_2$ (ps) | $t_{avg}$ (ps) |
|---|---|---|---|---|---|---|
| Fig. 2C | 7.56 | 0.1 | 36.6 | 0.9 | 734.3 | 252.66 |
| Fig. 2D (low power) | 4.70 | 0.05 | 35.4 | 0.95 | 544.2 | 316.64 |
| Fig. 2D (high power) | 6.50 | 0.05 | 35.4 | 0.95 | 544.2 | 316.64 |

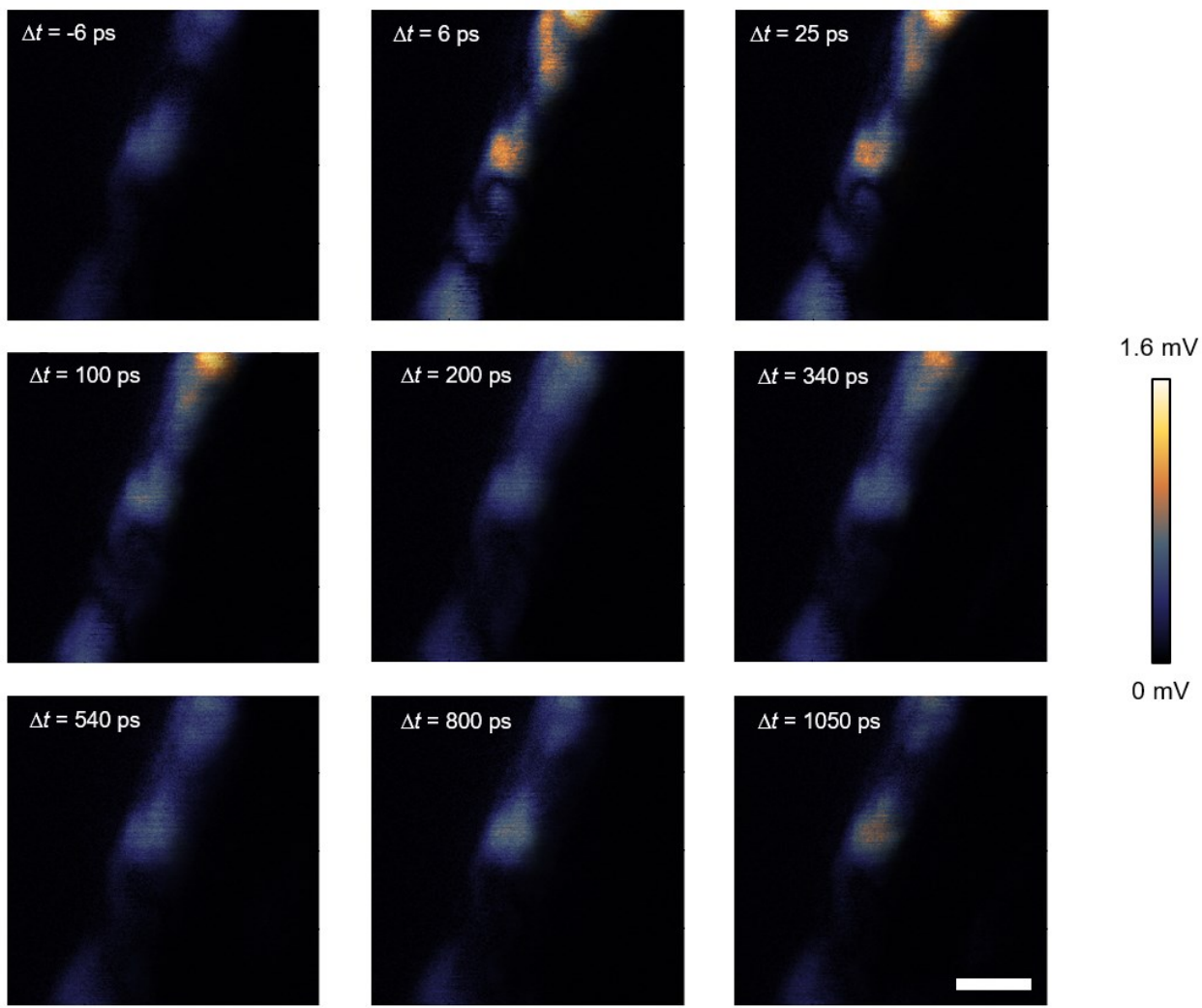

**Fig. S6. Time-resolved s-SNOM imaging of a nonuniform silicon nanostructure.** Scale bar: 2μm.